\documentclass[11pt,twoside]{article}
\usepackage{./asp2014}
\aspSuppressVolSlug
\resetcounters

\begin{document}

\title{Defocused Observations of Selected Exoplanet Transits with T100 in T\"{U}B\.{I}TAK National Observatory of Turkey (TUG)}
\author{\"Ozg\"ur Ba\c{s}t\"urk,$^{1,2}$ Tobias C. Hinse,$^3$ \.{I}brahim \"Ozavc{\i},$^{1,2}$ Onur Y\"or\"uko\u{g}lu,$^{1,2}$ and Selim O. Selam$^{1,2}$}
\affil{$^1$Ankara University, Faculty of Science, Dept. of Astronomy and Space Sciences, Tando\u{g}an, TR-06100, Ankara, Turkey; \email{obasturk@ankara.edu.tr}}
\affil{$^2$Ankara University Astronomy and Space Sciences Research \& Application Center, \.{I}ncek, TR-06837, Ankara, Turkey}
\affil{$^3$Korea Astronomy and Space Science Institute, 776 Daedukdae-ro, Yuseong-gu, 305-348 Daejeon, Republic of Korea}

\paperauthor{\"Ozg\"ur Ba\c{s}t\"urk}{obasturk@ankara.edu.tr}{ }{Ankara University}{Dept. of Astronomy and Space Sciences}{Ankara}{ }{TR-06100}{Turkey}
\paperauthor{Tobias C. Hinse}{tchinse@gmail.com}{ }{Korea Astronomy and Space Science Institute}{ }{Daejeon}{ }{ }{Republic of Korea}
\paperauthor{\.{I}brahim \"Ozavc{\i}}{ozavci@gmail.com}{ }{Ankara University}{Dept. of Astronomy and Space Sciences}{Ankara}{ }{TR-06100}{Turkey}
\paperauthor{Onur Y\"{o}r\"{u}ko\u{g}lu}{yorukoglu@gmail.com}{ }{Ankara University}{Dept. of Astronomy and Space Sciences}{Ankara}{ }{TR-06100}{Turkey}
\paperauthor{Selim O. Selam}{Selim.Selam@science.ankara.edu.tr}{ }{Ankara University}{Dept. of Astronomy and Space Sciences}{Ankara}{ }{TR-06100}{Turkey}

\begin{abstract}
It is crucial to determine masses and radii of extrasolar planets with high precision to have constraints on their chemical composition, internal structure and thereby their formation and evolution. In order to achieve this goal, we apply the defocus technique in the observations of selected planetary systems with the 1 m Turkish telescope T100 in T\"{U}B\.{I}TAK National Observatory (TUG). With this contribution, we aim to present preliminary analyses of transit light curves of the selected exoplanets KELT-3b, HAT-P-10b/WASP-11b, HAT-P-20b, and HAT-P-22b, observed with this technique using T100. 
\end{abstract}

\section{Introduction}
\label{sec:introduction}
Transiting Extrasolar Planets (TEPs) are key objects in determining physical (masses and radii) and atmospheric properties with high accuracy when their spectroscopic and photometric observations are combined. More than 1000 TEPs have been discovered so far in over 600 planetary systems\footnote{The Extrasolar Planets Encyclopedia: http://exoplanet.eu}, many of which do not have high-quality follow-up light curves. Deriving absolute parameters of an exoplanet is a difficult task, which relies on high quality photometric and spectroscopic data. In order to achieve the desired photometric precision, a very high signal-to-noise ratio (hereafter SNR) is essential. Defocused observations of bright targets is a well established technique based on collecting more photons and controlling the systematic affects arising mainly from imperfections of the detector and atmospheric seeing conditions \citep{southworth09, basturk14}. We aim to present the transit light curves of four selected exoplanets obtained by heavily defocusing the 1 m Turkish telescope T100 located in TUG with this contribution. 

\section{Observations and Data Reduction}
\label{sec:observations}
We make use of the defocusing technique to achieve high SNR by distributing the Point Spread Function (hereafter PSF) over many pixels which allow us to expose for longer durations, therefore diminish Poisson noise since the large PSFs are insensitive to focus or seeing changes \citep{southworth09}. The systemmatic errors brought about by flat-fielding is also reduced by two orders of magnitude with this technique.\\ 

We performed our observations with the 1 meter Turkish telescope, T100,  using an R$_{c}$ filter in close accordance with the Johnson-Cousins photometric system. The weather conditions were suboptimal in the observations of HATNet project planets while we had extremely stable conditions during the observations of KELT-3b on the night of February 18, 2014 letting us reach one of the best precision values ever acheived with this telescope in spite of a $\sim$~90\% waning moon's phase. We give a log of our observations in Table~\ref{tab:table1}.\\  

\begin{table}[t]
\small
\begin{center}
\caption{The log of the observations presented in this work}
\label{tab:table1}
\begin{tabular}{llccccc}
\hline\hline
Planet & Date  & St.Time & End.Time  & Exp.Time & \# of Pts. & Scatter \\
      &  & (UT) & (UT) & (s) &   & (mmags) \\
\hline
HAT-P-10b & 2013-01-15 & 16:28  & 20:21 & 120 & 88 &  0.89 \\
HAT-P-20b & 2014-02-19 & 17:41  & 21:32 & 90 & 91 & 0.78 \\
HAT-P-22b & 2014-02-17 & 22:58 & 03:37 & 150 & 79 & 1.42 \\
KELT-3b & 2014-02-18 & 17:39 & 03:20 & 150 & 161 & 0.58 \\
\hline\hline
\end{tabular}
\end{center}
\end{table}

We used a specific IDL code for decoused observations developed by \citet{southworth09} for all bias, dark, flat corrections, alignment of the images, and differential photometry with respect to a synthetic comparison star, which was formed by weighted flux summation of carefully selected stars to form an ensemble.

\section{Analysis}
\label{sec:analysis}
\begin{figure}
\centerline{\includegraphics[width=12cm,keepaspectratio]{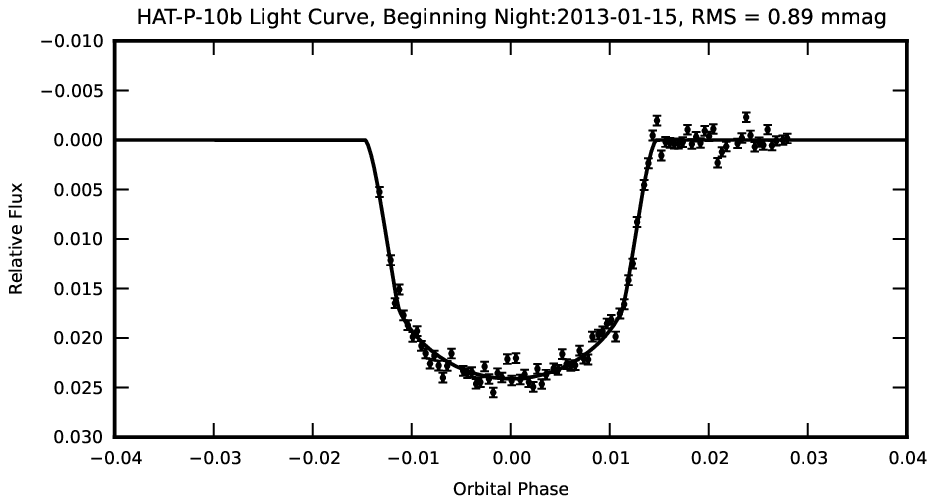}}
\caption{Transit observation of HAT-P-10b in defocused mode with T100}
\label{fig:fig1}
\end{figure}

\begin{figure}
\centerline{\includegraphics[width=12cm,keepaspectratio]{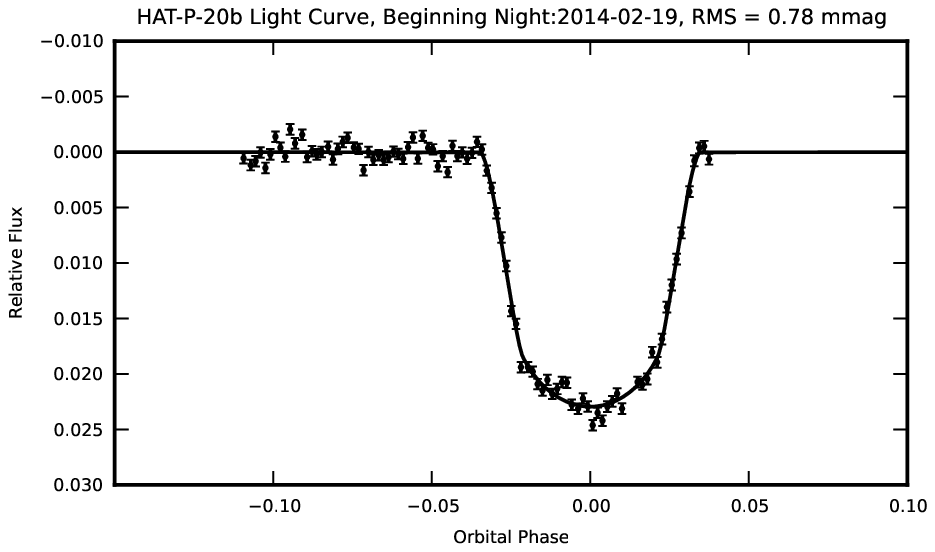}}
\caption{Transit observation of HAT-P-20b in defocused mode with T100}
\label{fig:fig2}
\end{figure}

\begin{figure}
\centerline{\includegraphics[width=12cm,keepaspectratio]{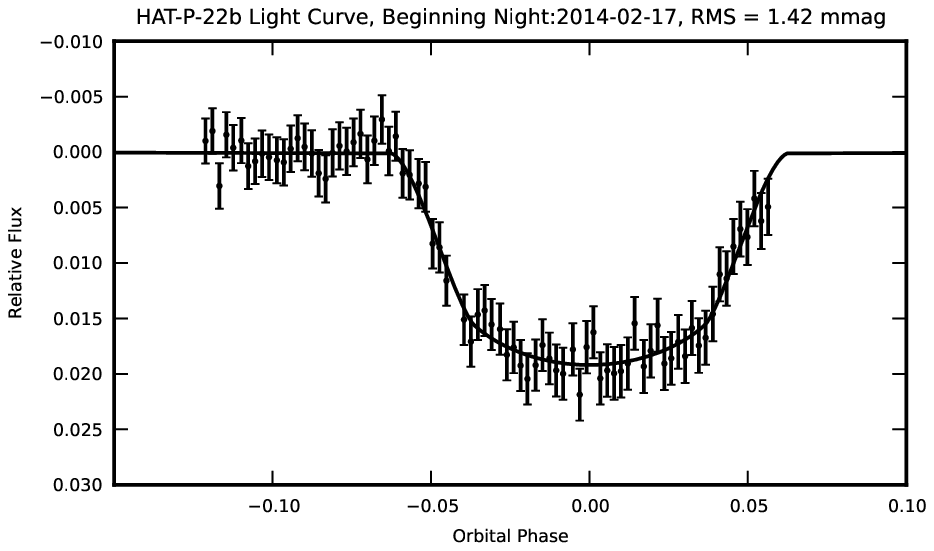}}
\caption{Transit observation of HAT-P-22b in defocused mode with T100}
\label{fig:fig3}
\end{figure}

\begin{figure}
\centerline{\includegraphics[width=12cm,keepaspectratio]{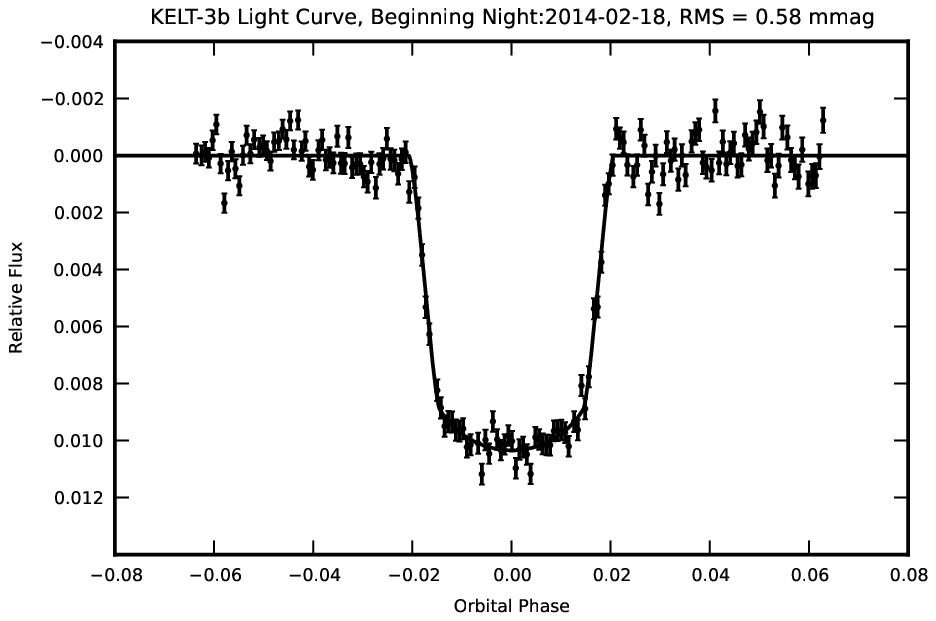}}
\caption{Transit observation of KELT-3b in defocused mode with T100}
\label{fig:fig4}
\end{figure}

The light curves were analysed with the JKTEBOP (version 34) code \citep{southworth08}, a modified version of EBOP \citep{popper81} for modeling transit light curves of exoplanets. We adjusted the inclination (i) parameter of the orbit, the sum (r$_{p}$ + r$_{s}$) and the ratio of the fractional radii (k = r$_{p}$ / r$_{s}$), the time of minimum of the light curve, while fixing the orbital periods to the values in Table~\ref{tab:table2} during our solutions. We used linear limb darkening law and fitted for the linear coefficient while holding the nonlinear coefficient fixed at the theoretical value. The best fits for the light curves of the selected exoplanets for this work are given in Figs.~\ref{fig:fig1},~\ref{fig:fig2},~\ref{fig:fig3}, and~\ref{fig:fig4}.

\begin{table}[t]
\small
\begin{center}
\caption{Information for targets in Figures~\ref{fig:fig1},~\ref{fig:fig2},~\ref{fig:fig3}, and~\ref{fig:fig4}}
\label{tab:table2}
\begin{tabular}{lccccc}
\hline\hline
Planet & Period  & Epoch & V & Depth & Duration \\
      & (days) &  & (mag)  & (mag) & (min)\\
\hline
HAT-P-10b & 3.722469 & 2454729.90631 & 11.89 & 0.0254 & 159\\
HAT-P-20b & 2.875317 & 2455080.92661 & 11.34 & 0.0204 & 111\\
HAT-P-22b & 3.212220 & 2454930.22001 &  9.73 & 0.0119 & 172\\
KELT-3b & 2.703480 & 2456023.459 & 9.86 & 0.0098 & 197\\
\hline\hline
\end{tabular}
\end{center}
\end{table}

\section{Results \& Conclusion}
\label{sec:results}
We summarize our results in Table~\ref{tab:table3}. Ratio of the radius of the planet to that of the star (k = r$_{p}$ / r$_{s}$) is the most important parameter one can deduce from a light curve. The high precision that we achieved in the measurements of the parameter is mostly due to the quality of the observations. However, we should stress that more light curves are needed to perform a rigorous analysis and the errors given in this study are mostly the formal errors obtained from the covariance matrix of the fitting (Levenberg-Marquardt) algorithm. Hence it will not be adequate to compare the resultant parameters with the previously published values. We have also computed the mid-transit times end their errors for these transits in Table~\ref{tab:table3} by fitting parabolas to the flat bottom sections of the light curves.\\

\begin{table}[t]
\small
\begin{center}
\caption{Parameters of the fits to the light curves of the selected exoplanets}
\label{tab:table3}
\medskip
\begin{tabular}{lccccc}
\hline
Planet & r$_{s}$ + r$_{p}$  & k = r$_{p}$ / r$_{s}$ & i ($^{\circ}$) & T0 (+2400000) \\
\hline
HAT-P-10b & 0.0924 $\pm$ 0.0003 & 0.1345 $\pm$ 0.0005 &  89.8 $\pm$ 0.1 & 56308.237798 $\pm$ 0.000107 \\
HAT-P-20b &  0.2558 $\pm$ 0.0044 &  0.1393 $\pm$ 0.0010 & 81.9 $\pm$ 0.1  & 56708.356256 $\pm$ 0.000088  \\
HAT-P-22b & 0.4637 $\pm$ 0.0394 & 0.1294 $\pm$ 0.0049 & 73.6 $\pm$ 3.7 & 56706.589360 $\pm$ 0.000900 \\
KELT-3b  & 0.1619 $\pm$ 0.0047 & 0.0957 $\pm$ 0.0006 & 84.6 $\pm$ 0.4 & 56707.438255 $\pm$ 0.000207 \\
\hline
\end{tabular}
\end{center}
\end{table}

Our results show that T100 has a decent potential for follow-up transit observations of exoplanet transits, with the defocusing method in particular, giving a sub-milimagnitude photometric precision. We aim to accumulate additional data in the future and we are willing to share the data presented here with other researchers for them to carry out a more detailed/robust analysis.

\acknowledgements
We thank to  T\"{U}B\.{I}TAK for the partial support in using T100 telescope with project number 12CT100-378. Authors from Ankara University also acknowledge the support by the research fund of Ankara University (BAP) through the project 13B4240006. TCH is supported via the Korea Young Scientist Research Fellowship (KRCF) carried out at the Korea Astronomy and Space Science Institute (KASI) and acknowledges financial support under KASI grant number 2013-9-400-00.



\end{document}